\documentclass[prb,twocolumn,superscriptaddress,oneside, floatfix]{revtex4}%
\usepackage{amssymb}
\usepackage{amsfonts}
\usepackage{graphicx}
\usepackage{epsfig}
\usepackage{amsmath}
\begin{document}

\title{Reply to Comment on ``High-field studies of superconducting fluctuations in high-T$_c$ cuprates: Evidence for a small gap distinct from the large pseudogap'' by M.V. Ramallo et al.}
\author{F. Rullier-Albenque}
\email{florence.albenque-rullier@cea.fr}
\affiliation{Service de Physique de l'Etat Condens\'e, Orme des Merisiers, CEA Saclay
(CNRS URA 2464), 91191 Gif sur Yvette cedex, France}
\author{H. Alloul}
\affiliation{Laboratoire de Physique des Solides, CNRS UMR 8502, Universit\'e Paris Sud 11,
91405 Orsay, France}
\author{G. Rikken}
\affiliation{Laboratoire National des Champs Magn\'{e}tiques Intenses, UPR 3228, CNRS-UJF-UPS-INSA, 31400 Toulouse, France}
\date{\today}

\begin{abstract}
The experimental investigations done in our paper Phys.Rev.B\textbf{84},014522(2011) allowed us to establish that the superconducting fluctuations (SCF) always die out sharply with increasing $T$. But contrary to the claim done in the comment of Ramallo \textit{et al.}, this sharp cutoff of SCF measured in YBa$_2$Cu$_3$O$_{6+x}$ depends on hole doping and/or disorder. So our data cannot be used to claim for a universality of the extended gaussian Ginzburg Landau theory proposed by the authors of the comment. Furthermore, to explain quantitatively our data near optimal doping using this model they need to consider that fluctuations in the two CuO$_2$ planes of a bilayer are totally decoupled, which is not physically well justified. On the contrary a consistent interpretation of all our data (paraconductivity, Nernst effect and magnetoresistance) has been done by considering that the coupling between the two layers of the unit cell is dominant at least up to 1.1$T_c$. 
\end{abstract}

\maketitle

First let us point out that the comment of Ramallo \textit{et al.} about our paper \cite{FRA-PRB2011}
concerns mainly the theoretical interpretation of the fast decrease of the superconducting fluctuations 
(section VII-A) that has been detected very early at optimal doping (see ref.55-57 of 
ref.\cite{FRA-PRB2011}). The corresponding onset of superconductivity with decreasing temperature has 
been quite diversely appreciated during the last decade with different experimental techniques, especially 
in the underdoped regime. The experimental studies performed so far (see ref.6,7,8,10,11 of 
ref.\cite{FRA-PRB2011}), including those listed in ref.13 
of the comment, have attempted determinations of the onset of superconducting fluctuations (SCF) but 
have most often neglected to perform analyses of the magnitude of the measured physical quantities. 

As noticed by the authors of the comment, we have performed in our work a very thorough determination 
of the normal state transport properties using high magnetic field experiments in order to precisely 
extract the SCF contribution to the conductivity. This method which we had established on an 
underdoped sample \cite{RA-HF}  has been extended by the study of samples around optimal doping 
\cite{Alloul-EPL2010}. Following wishes of some members of the scientific community we have done in 
ref.\cite{FRA-PRB2011} a detailed presentation of the data obtained and performed there, a more
extended quantitative analysis of this kind of data than performed so far. These results 
therefore cannot be compared with previous data that have been analysed using temperature 
extrapolations which are highly questionable especially for underdoped samples as detailed in 
the introduction of our paper.  In particular, we could have mentioned there the paper of 
Curras et al \cite{Curras} on La$_{2-x}$Sr$_x$CuO${_4}$ (LSCO) thin films which is heavily quoted in the comment 
as particularly representative of such a data analysis which cannot be considered as reliable. 
Thus our results \textit{cannot be considered as 
a useful confirmation} of such experimental studies. 

One important point we have shown in great detail in our paper is that for all dopings, in pure 
and disordered YBa$_2$Cu$_2$O$_{6+x}$ (YBCO) samples the fluctuations die sharply versus field and temperature as shown in 
Fig.17, 18, 26 and 27 of ref.\cite{FRA-PRB2011}. This point is important as it means that 
the determinations of the onset of SCF at $T^{\prime}_c$ or $H^{\prime}_c$ are well defined 
and not much dependent on the sensitivity threshold used. 

Concerning the quantitative analysis of our data near optimal doping, we first considered in 
ref.\cite{FRA-PRB2011} the behaviour found above $T_c$ up to $\epsilon \sim 0.1$: in this temperature 
range a good agreement is found with the Lawrence-Doniach (LD) approach of SCF in quasi two-dimensional systems. We took 
into account the bilayering effect present in YBCO by considering that the two CuO$_2$ layers of the 
unit cell are coupled, that is $s = c=11.7$\AA ~with $c$ the c-axis parameter, as has been done by
most authors so far (see for instance ref. 21, 24, 26, 27, 29 of ref.\cite{FRA-PRB2011}).  

This choice of $s=c$ allows us to explain quite consistently our experimental data for 
the paraconductivity, the Nernst effect and the magnetoconductance in the temperature range corresponding 
to $\epsilon$ between ~0.01 and ~0.1. We could also account there for our magnetoconductance data in the framework of 
the Aslamazov-Larkin theory. Moreover, as indicated in our paper, the value of $H_{c2}(0)$ obtained 
for the optimally doped sample is found to correspond to a coherence length $\xi(0)=1.33$nm, in 
excellent agreement with that found independently from the comparison between the paraconductivity and 
the Nernst effect, as shown in Fig.19 of our paper using the relationship Eq.13 in which the parameter $s$ has been eliminated. Therefore, all the analyses performed in ref.\cite{FRA-PRB2011} appear to be consistent and quite compatible with taking into account the coupling within the bilayer.

The criticism of the comment is threefold. 

(i)	The authors consider that Vidal \textit{et al.} \cite{Vidal} have established theoretically 
that the cutoff is a universal extension of the Gaussian Ginzburg Landau (GGL) theory, 
which should apply to any superconductor. 
Then by adapting the fit parameters, they consider that our results do prove the validity of this theory 
for all dopings above 0.1 and by extension for all the cuprate families. On the contrary we have 
been extremely careful to show on experimental grounds that the apparent universality near optimal 
doping does not apply at all for underdoped samples and for situations in which disorder prevails 
(see fig. 17 and 18 of ref.\cite{FRA-PRB2011}). Ramallo \textit{et al.} refute the validity of our data taken 
on the most underdoped YBCO sample on the basis that these can be contaminated by some distribution in $T_c$. 
On the contrary let us point out that the underdoped YBCO with $T_c=57.1$K used in our study has a 
narrow transition width of 0.6K, which signals a very good homogeneity for this sample. More generally, 
it has been shown by NMR measurements \cite{Bobroff-PRL2002} that YBCO samples near this composition 
are much more homogeneous than any other cuprate families. On the contrary LSCO is by far one of the cuprate 
families displaying most inhomogeneities, especially in the underdoped regime \cite{FRA-EPL-MIT}. 
One might naturally wonder why the data of Curras et al. \cite{Curras} taken on LSCO films with
transition widths as large as 5K are found to agree with Vidal's model. 

Let us also mention here that the same type of analysis on similar underdoped LSCO samples have 
been performed by B. Leridon \textit{et al.} \cite{Leridon-PRB2007} who used high magnetic fields to 
suppress superconductivity and determine the normal state resistivity. There the fluctuation conductivity
could be very well accounted for by the 2D Aslamazov-Larkin formula up to $\epsilon \sim 0.2$ and then 
bent downwards and vanished only for $\epsilon \sim ~0.9$. But these results have been also rejected by 
Ramallo \textit{et al.} (see ref.5 of the comment). Does not that demonstrate that an appropriate choice 
of fitting parameters to extract the SCF permits to fit the data? 
In any case, it is not serious to qualify the validity of experimental results by 
the ability to adapt the parameters to reproduce the results of a single given theoretical model!

(ii) We do not grasp the physical validity of the arguments taken by Vidal \textit{et al.} to estimate 
that the critical value of the cutoff must always corresponds to $\epsilon^c=0.5$, 
and we have seen no argument in this comment explaining why samples with reduced $T_c$ 
either by doping or disorder would display different values of $\epsilon^c$ extending from 0.3 to 1 
as found in our data or in others. In the case of disordered films values up 
to 2 have even been measured \cite{Pourret-NPhys2006}. So we consider that our data cannot be considered in 
any manner as a support for a universal validity of the phenomenological expression proposed by
Vidal \textit{et al.}.

(iii) Contrary to LSCO compounds for which $s$ is simply the distance between the CuO$_2$ planes, the case 
of YBCO with the existence of bi-layers introduces some complications. If we limit the quantitative
considerations to the pure samples near optimal doping, the data are not fitted correctly by their
phenomenological expression (Eq.9 of ref.10 of the comment) with $s=c$ for $\epsilon <0.1$, 
as noticed by the authors of the comment themselves (see their Fig.1)\cite{footnote}. To fit correctly 
our data, Ramallo \textit{et al.} need to take an interlayer periodicity $s=c/2$, that is 
to neglect totally the coupling within the bi-layer. Contrary to their claim, using the value $s=c/2$ is 
not an assumption well established in the community. References 9 and 15 quoted in the comment
are exclusively from their own work. One can notice in Fig.1 of the comment that their fit with this value 
of $s$ does not reach the LD limit even for the lowest $\epsilon$ value considered in our work, 
showing surprisingly that the incidence of the cutoff would be already at play for $T\sim1.01T_c$. 

On physical grounds, as the coupling within the bi-layer is larger than that between bilayers, one 
naturally expects the LD correction due to the coupling between bilayers to be dominant near $T_c$.  
While in our analysis we only consider the progressive inter bilayer decoupling with increasing $T$, 
the decoupling of the planes of the bilayer considered by Ramallo \textit{et al.} should only be effective
at higher $T$. This will roughly occur when $2\xi_c(0)/\epsilon^{1/2}$, with $\xi_c(0)$
the transverse coherence length, becomes comparable to the spacing ($\sim 3.4$\AA) between the planes 
of the bi-layer, that is $\epsilon \simeq 0.1$ for $\xi_c(0)=0.9$\AA. Thus from $\epsilon=0.1$ down 
to $\epsilon \sim 0.02$ one would expect to enter a regime where the bi-layers are coupled before 
reaching a 3D behavior near $T_c$. In such a case $s=c$ should apply between 0.02 and 0.1.  

So the phenomenological expression proposed by Ramallo \textit{et al.} can be made to fit some 
of our data when appropriate parameter choices are done. But this does not allow to conclude for its 
validity or, even less, that it applies for all the cuprates. This justifies our presentation, which only
insists on the existence of a sharp cutoff for all the samples with the same evolution of the 
fluctuation conductivity in function of temperature and magnetic fields. Our work definitely allowed us 
to  demonstrate that Gaussian fluctuations do explain fully the SCF around and above optimal doping. 
We  believe that more experimental and theoretical work is still needed to  establish the physical origin 
of the onset of pairing.


\begin{thebibliography}{99}

\bibitem{FRA-PRB2011} F.Rullier-Albenque, H. Alloul, G. Rikken, Phys. Rev. B \textbf{84}, 014522 (2011).
 
\bibitem{RA-HF} F. Rullier-Albenque, H. Alloul, C. Proust, D. Colson, and A. Forget, Phys. Rev. Lett. \textbf{99}, 027003 (2007).

\bibitem{Alloul-EPL2010} H. Alloul, F. Rullier-Albenque, B. Vignolle, D. Colson, and A. Forget, Europhys. Lett. \textbf{91}, 37005 (2010).

\bibitem{Curras} S.R. Curras et al. Phys. Rev. B \textbf{68}, 094501 (2003).

\bibitem{Vidal} F. Vidal et al., Europhys. Lett. \textbf{59}, 754 (2002). 

\bibitem{Bobroff-PRL2002} J. Bobroff et al., Phys. Rev. Lett. \textbf{89}, 157002 (2002). 

\bibitem{FRA-EPL-MIT} F. Rullier-Albenque, H. Alloul, F. Balakirev, C. Proust, Europhys. 
Lett. \textbf{81}, 37008 (2008).

\bibitem{Leridon-PRB2007} B. Leridon, J. Vanacken, T. Wambecq, and V. V. Moshchalkov, 
Phys. Rev. B textbf{76}, 012503 (2007).

\bibitem{Pourret-NPhys2006} A. Pourret, H. Aubin, J. Lesueur, C. A. Marrache-kikuchi, L. Berg\'e, L. Dumoulin, and K. Behnia, Nat Phys. \textbf{2}, 683 (2006).

\bibitem{footnote} One should be aware that logarithmic plots such as those displayed in Fig.1 of the comment
do artificially mask the discrepancies. There, the fitting curve differs from the experimental data by 
at least $\sim25$\%, well beyond the actual experimental error bars. This explains the sentence used in 
ref.\cite{FRA-PRB2011}: ``this equation privileges the cutoff behavior and does not reproduce the 
low-$T$ regime where we have established the validity of the GL approach''. 




\end{thebibliography}
\end{document}